\documentclass[sigconf]{acmart}
\AtBeginDocument{%
  }

\begin{document}
\title{The Poisoned Chalice of LLM Evaluation Report}

\author{Jonathan Katzy}
\email{J.B.Katzy@TUDelft.nl}
\orcid{0009-0005-9574-2414}
\authornote{Main contact}
\affiliation{%
  \institution{Delft University of Technology}
  \city{Delft}
  \country{The Netherlands}
}
\author{Ali Al-Kaswan}
\email{A.Al-Kaswan@TUDelft.nl}
\orcid{0000-0001-7338-2044}
\affiliation{%
  \institution{Delft University of Technology}
  \city{Delft}
  \country{The Netherlands}
}
\author{Razvan Mihai Popescu}
\email{R.M.Popescu@TUDelft.nl}
\orcid{0009-0003-6251-770X}
\affiliation{%
  \institution{Delft University of Technology}
  \city{Delft}
  \country{The Netherlands}
}
\author{Zhou Yang}
\email{ZY25@UAlberta.ca}
\orcid{0000-0001-5938-1918}
\affiliation{%
  \institution{University of Alberta, \\Alberta Machine Intelligence Institute}
  \city{Alberta}
  \country{Canada}
}

\acmArticleType{Research}
\copyrightyear{2026}
\acmYear{2026}
\setcopyright{cc}
\setcctype{by}
\acmConference[FSE Companion '26]{34th ACM Joint European Software Engineering Conference and Symposium on the Foundations of Software Engineering}{July 05--09, 2026}{Montreal, QC, Canada}
\acmBooktitle{34th ACM Joint European Software Engineering Conference and Symposium on the Foundations of Software Engineering (FSE Companion '26), July 05--09, 2026, Montreal, QC, Canada}
\acmDOI{10.1145/3803437.3807733}
\acmISBN{979-8-4007-2636-1/2026/07}



\begin{abstract}
Large language models are increasingly used to evaluate and support software engineering tasks, yet the validity of these evaluations is often undermined by uncertainty about whether benchmark instances were seen during pretraining. This can lead to data contamination, which may inflate performance and result in misleading conclusions about model capability. Despite this, the training corpora of many modern models are only partially disclosed, making direct decontamination infeasible. This creates a need for practical methods that can detect a large language models' prior exposure to training data without access to the full training corpus.

To address this challenge, we organize the first Poisoned Chalice of LLM Evaluation Competition, co-located with the FSE-AIWare 2026 Competition Track. The competition frames contamination detection as a white-box membership inference task on source code and provides participants with curated datasets, target models, baseline attacks, and a final evaluation on a held-out model and dataset. This design encourages methods that generalize beyond superficial dataset artifacts and beyond a single training setting.

This paper reports the setup and results of the competition. More broadly, the competition aims to catalyze the community around trustworthy LLM evaluation for software engineering.

\end{abstract}
\maketitle
\section{Introduction}
Large Language Models (LLMs) have become central to modern software engineering research and practice. They have been applied to a wide range of tasks, including code generation, code completion, code summarization, test generation, vulnerability detection, and program repair~\cite{hou2024large, yang2025ecosystem}. Their increasing effectiveness has also led to widespread adoption in developer tooling, where they are used to accelerate programming and support day-to-day software development activities.

At the same time, the rapid progress of code-capable LLMs has made rigorous evaluation increasingly difficult. A central challenge is that LLMs are known to memorize portions of their training data and, in some cases, reproduce them verbatim or near-verbatim~\cite{al2024traces, carlini2021extracting, yang2024unveiling}. As a result, strong performance on a benchmark may not necessarily reflect robust generalization; instead, it may partially arise from prior exposure to benchmark instances during training. This issue, commonly referred to as \emph{data contamination}, threatens the validity of downstream evaluation and can lead to overly optimistic conclusions about model capability~\cite{al2023abuse}.

In open-data settings, contamination can sometimes be mitigated by decontaminating the evaluation set against known training corpora. While licenses can be a weak pre-filter, all large open datasets contain significant amounts of copied code, even across licenses~\cite{katzy2024exploratory}. We have previously addressed this and constructed \emph{The Heap}, a dataset designed for evaluating language models on code tasks~\cite{katzy2025heap}. However, even in such cases, decontamination can be computationally expensive due to the scale of contemporary training corpora. For example, \emph{The Stack v2}, a widely used corpus for code LLM training, exceeds 90TB in size~\cite{lozhkov2024starcoder2stackv2}. More importantly, this strategy is not applicable to closed or partially closed models whose training data are not publicly disclosed~\cite{al2024traces}. For such models, direct deduplication against the training corpus is impossible.

This limitation motivates the need for alternative methods that can estimate whether a specific artifact was likely seen during training. A natural direction is provided by \emph{membership inference attacks} (MIAs), which aim to infer whether a given sample was part of a model's training data by exploiting differences in model behavior between members and non-members~\cite{carlini2022membership}. In the context of code LLM evaluation, MIAs offer a promising mechanism for assessing potential contamination without requiring direct access to the underlying pretraining corpus.

Despite this promise, membership inference remains an open problem. Existing approaches often exhibit limited reliability in realistic settings, and evidence on source code is still limited~\cite{hou2024large}. Moreover, much of the prior literature evaluates MIAs on short textual fragments or narrowly scoped instances, whereas practical contamination assessment for software engineering benchmarks often requires reasoning at the level of complete files.

To advance research in this area, we organized the first edition of the \emph{Poisoned Chalice of LLM Evaluation Competition}, co-located with the FSE-AIWare 2026 Competition Track. The competition focuses on white-box membership inference for code LLMs. Participants are given development datasets containing member and non-member code files, along with open-weight target models, and are asked to design methods that distinguish between files that were included in training and files that were not.

To encourage methods that go beyond superficial dataset artifacts, the competition datasets were curated to remove comparatively easy cases, for example those distinguishable primarily through temporal or metadata shifts. Final evaluation is performed on a held-out model and a held-out dataset not accessible during development, thereby emphasizing generalization rather than overfitting to a single model or collection of files.

\section{Evaluation}
The central objective of the competition is to assess how effectively a method detects whether a code file has been used in the training of a large language model. This objective reflects the broader goal of supporting reliable code LLM evaluations, especially in settings where direct access to training corpora is unavailable. We therefore design the evaluation to measure not only predictive performance on a known target model, but also the extent to which submitted approaches generalize across models and datasets.

To support this objective, we provide competitors with two labeled development datasets. One dataset contains code files that are included in the training data of StarCoder2, while the other contains code files that are neither exact nor near duplicates of files used during training. These datasets allow participants to develop and calibrate their membership inference methods under controlled conditions. In addition, we maintain a held-out evaluation dataset with similar characteristics but different files, which we use for the final assessment of submissions. To evaluate whether approaches generalize beyond a single target model, we also include a validation model together with a corresponding held-out dataset following the same member/non-member construction principles.

\subsection{Model Selection}
To choose an appropriate dataset for this competition, several key considerations had to be taken into account. First, we required a dataset–model pair for which it is known which data were included in training and which were excluded. Although many models are described as open source, this openness is often limited to the release of model weights, while the training data remain undisclosed. Consequently, the set of suitable models is restricted to those trained on public datasets. The main candidates are StarCoder2~\cite{lozhkov2024starcoder2stackv2}, SmolLM2~\cite{allal2025smollm2smolgoesbig}, and Mellum~\cite{pavlichenko2025mellum}. As we are interested in scalability and generalizability of approaches in combination with performance we selected StarCoder2 3B and 7B for the competition. Mellum-4B was retained as a held-out model.

\begin{figure*}[tb]
    \centering
    \includegraphics[width=1\linewidth]{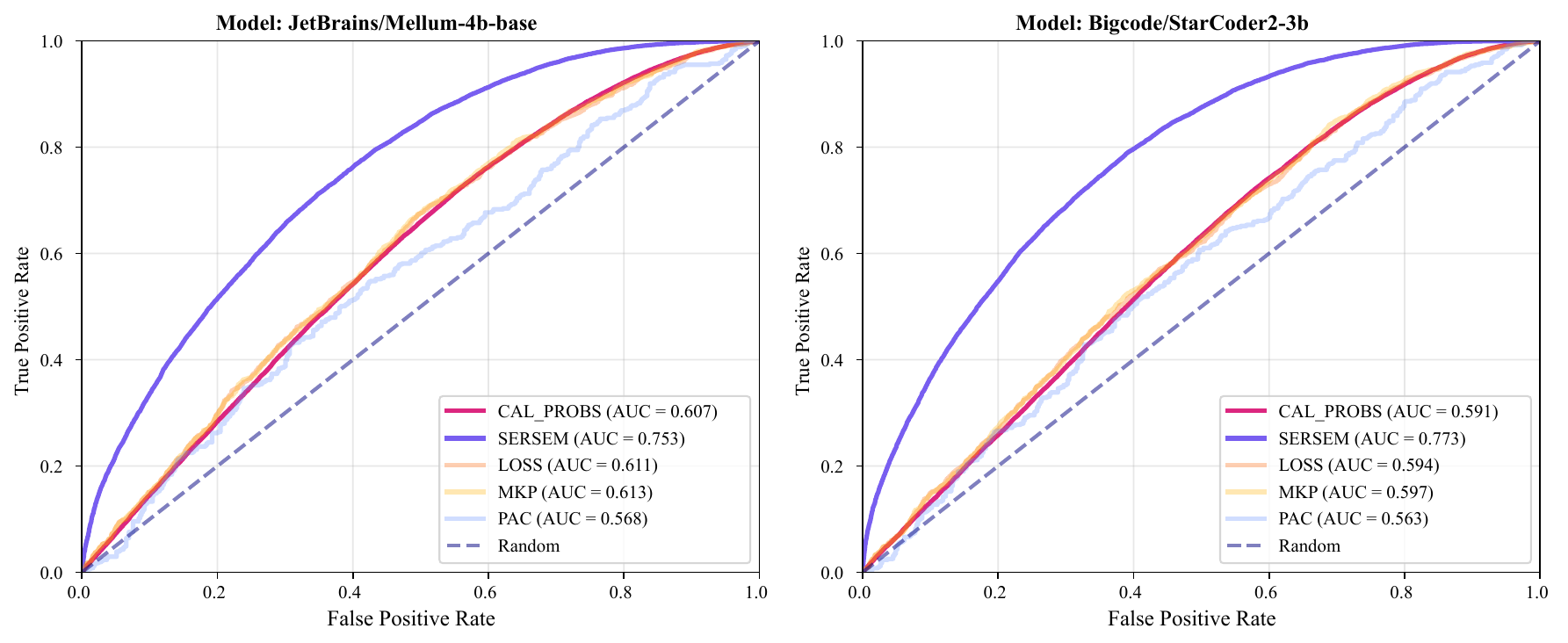}
    \caption{ROC curves for all evaluated membership inference methods on the final test sets for \texttt{Mellum} and \texttt{StarCoder2-3B}}
    \Description{Two side-by-side ROC plots comparing membership inference methods on Mellum-4B and StarCoder2-3B. Each plot includes curves for CAL\_PROBS, LOSS, MKP, PAC, SERSEM, and a dashed random baseline. In both plots, the SERSEM curve lies clearly above the others and has the highest area under the curve, with AUC values of $0.753$ on Mellum-4B and $0.773$ on StarCoder2-3B. The remaining methods cluster closer to the diagonal, with AUC values between about $0.56$ and $0.61$.}
    \label{fig:roc}
\end{figure*}

\subsection{Dataset Curation}
To construct the dataset, we required two data sources: one that had been seen during training by the model and one that had definitely not been seen. To achieve this, we used both \emph{The Stack Edu}~\cite{allal2025smollm2smolgoesbig} and The Heap~\cite{katzy2025heap} as data sources. We selected The Stack Edu as the source of seen data because it is a subset of \emph{The Stack v2} and was therefore included in the training data of Mellum and the StarCoder2 models. This also makes the resulting dataset suitable for experiments with SmolLM, although that model was outside the scope of this competition. We selected The Heap as the source of unseen data because it has been deduplicated against \emph{The Stack v2}, which is a superset of \emph{The Stack Edu}. From The Heap, we select files based on the near duplicates column, which uses locality sensitive hashes to also remove files that had only minor changes made to them, resulting in a more diverse dataset. From these datasets, we selected a subset of languages for the competition, namely Go, Java, Python, Ruby, and Rust, to provide a broader overview of the generalizability of the submitted approaches across languages.

This process produced a large pool of potential member and non-member examples. However, small changes in functions, dates, or lexical choices in code can introduce a distribution shift between the member and non-member sets. To mitigate this effect, we trained a bag-of-words classifier to distinguish between the two sets and retained only the files that were misclassified by the classifier~\cite{meeus2025sok}. This resulted in member and non-member sets that could not be distinguished simply by the presence of specific keywords. From this curated dataset, participants received two subsets: a training split and a test split~\footnote{https://huggingface.co/datasets/AISE-TUDelft/Poisoned-Chalice}. We retained an additional unseen validation split that was not released to the participants.

\subsection{Evaluation Details}
We evaluate submissions using the Area Under the Receiver Operating Characteristic Curve (AUC-ROC). This metric is widely used in membership inference and binary classification settings because it captures the trade-off between the true positive rate and the false positive rate across all decision thresholds. We choose AUC-ROC instead of threshold-specific measures such as TPR@$x$FPR~\cite{al2023targeted}, because the community does not yet agree on a single acceptable false positive rate for this task. AUC-ROC therefore provides a threshold-independent and comparable measure of performance.

The final ranking reflects performance on held-out data that is not available to participants during development. This design reduces the risk that submissions overfit to the provided datasets or to artifacts specific to a single model. Since the competition aims to identify practically useful approaches for contamination detection, strong submissions need to perform well not only on the development setting, but also on unseen files and on a separate evaluation model.

We execute all submissions on our own hardware to ensure a uniform and reproducible evaluation environment. For this reason, each team submits a runnable replication package together with its method. We provide a dedicated description of the available hardware and execution environment on the competition website. After the submission deadline, we allow a short grace period during which we run all submitted systems and contact teams if minor packaging or execution issues require clarification or repair. This process helps ensure that all entries are evaluated fairly and under identical conditions.

\subsection{Baselines}
We provide three baseline membership inference attacks (MIAs) as reference implementations: \emph{Loss}, \emph{MinK\%Prob}, and \emph{PAC}. These baselines represent different strategies for inferring whether a sample belongs to the training data of a model.\footnote{Baselines: \url{https://github.com/AISE-TUDelft/PoisonedChalice}} \emph{Loss} uses the average token-level negative log-likelihood of a sample, relying on the intuition that training members tend to be assigned lower loss than non-members. \emph{MinK\%Prob} focuses only on the lowest-probability tokens in a file, under the assumption that rare or difficult tokens provide a stronger signal of non-membership than well-predicted boilerplate. Finally, \emph{PAC} compares the model's behavior on the original input and on perturbed variants, using this calibration step to better distinguish memorized samples from merely easy ones.

Each baseline produces a score that reflects the estimated likelihood that a given code file is a member of the training set. Together, these methods establish reference points for the competition and help participants position their approaches relative to existing techniques. We do not treat the baselines as exhaustive; rather, we include them to provide common starting points and to support a more interpretable comparison of submitted methods. For methodological details on the individual attacks, we refer the reader to the corresponding original publications~\cite{meeus2025sok}.

\section{Submitted Reports}
In this section we summarize the approaches submitted by participating teams. Since the competition encouraged diverse white-box membership inference strategies, the submitted reports reflect a range of design choices. Below, we provide concise overviews of the two submitted systems and their reported findings.

\subsection{SERSEM: Selective Entropy-Weighted Scoring for Membership Inference in Code Language Models}
\citeauthor{dikici2026sersem} propose \emph{SERSEM}~\cite{dikici2026sersem}, a structure-aware membership inference method that down-weights predictable syntactic boilerplate and emphasizes human-centric signals such as comments, long identifiers, string literals, formatting anomalies, and developer markers like \texttt{TODO}. Their approach combines weighted token-level scoring from output logits with probing of intermediate transformer activations, motivated by the idea that memorization signals may be more visible in hidden representations than in final predictions.

On a held-out portion of the training set, \emph{SERSEM} reports an overall AUC-ROC of $0.79$ on StarCoder2-3B and $0.79$ on StarCoder2-7B, outperforming the baselines across all five languages considered.

\subsection{Evaluating Signals for Membership Inference Attacks on Source Code}
\citeauthor{berndt2026evaluating} study simpler probability-based signals adapted to source code~\cite{berndt2026evaluating}. They evaluate several approaches, including average token log-probability, a slope-based signal over token probabilities, and Min-K\%++. Their best-performing method, \emph{CalibratedProbs}, adjusts average log-probability by token diversity, based on the intuition that repetitive code is easier to predict regardless of membership.

On a held-out set, they report an AUC-ROC of $0.66$ for \emph{CalibratedProbs} on StarCoder2-3B, improving over the strongest provided baselines by about $6$ percentage points. They also found that truncating the files to the first $200$ or $300$ tokens reduced performance.

\section{Evaluation Results}

\autoref{fig:roc} presents the ROC curves for all methods on \texttt{StarCoder2-3B} and the held-out model \texttt{Mellum}.\footnote{Due to runtime constraints, PAC was evaluated on 1{,}000 samples, whereas all other methods were evaluated on 5{,}000 samples.} Overall, the results show a clear separation between the strongest submitted method and the probability-based baselines. Across both models, \emph{SERSEM} achieves the highest AUC-ROC by a substantial margin, whereas the remaining methods cluster only slightly above random guessing. This indicates that the competition setting was challenging and that simple likelihood-based signals alone were insufficient to robustly distinguish member from non-member files on the curated evaluation data.

Across both models, \emph{SERSEM}~\cite{dikici2026sersem} achieves the strongest performance by a clear margin. On \texttt{StarCoder2-3B}, it reaches an AUC-ROC of $0.773$, compared with $0.597$ for \emph{MKP}, $0.594$ for \emph{LOSS}, $0.591$ for \emph{CAL\_PROBS}~\cite{berndt2026evaluating}, and $0.563$ for \emph{PAC}. On the held-out \texttt{Mellum}, \emph{SERSEM} again ranks first with an AUC-ROC of $0.753$, while the remaining methods achieve between $0.568$ and $0.613$.

Comparing the two submitted approaches, the final evaluation also reveals a generalization gap. While \emph{CalibratedProbs} improved over the provided baselines during participant-side experiments, its final AUC-ROC remains close to the other probability-based methods and below \emph{SERSEM}. In contrast, \emph{SERSEM} not only outperforms all reference baselines, but also achieves strong performance on the held-out Mellum model. This makes it the most effective submission in the competition and the clear winner.

\section{Conclusion and Closing Remarks}
The goal of this competition was to evaluate not only the performance of different MIAs but also to discuss their adaptability to real-world problems. Our results suggest that the limited generalizability and sensitivity to keyword drift identified in prior work~\cite{meeus2025sok} become less problematic when such confounds are explicitly addressed during dataset construction and evaluation design. However, to address issues such as data leakage and contamination, there are still two concerns. First, all attacks, including the baselines, are computationally complex. This is partially due to the need to run the inputs through an LLM, but also the need to run multiple LLM inferences for input perturbations can multiply this cost. 
SERSEM could be optimized for speed, for example by prioritizing a smaller number of early or intermediate layers instead of probing the full network.
Finally, there remains a practical trade-off between precision and recall. If the goal is to identify a subset of files that were almost certainly seen during training, one must operate at a very low false positive rate, which in practice means retaining only a small fraction of all positively scored samples. Conversely, removing as many seen samples as possible requires operating at a higher true positive rate, which also increases the false positive rate and causes genuinely unseen samples to be discarded.

\section{Acknowledgements}
We thank all the participants for their submissions, the program committee members, Ziyou Li and Roham Koohestani, who reviewed the submissions. The baseline implementations were developed by Cosmin Vasilescu, Ísak Jónsson, and Roham Koohestani as part of the Research Project course (CSE3000) at TU Delft.
\newpage

\bibliographystyle{ACM-Reference-Format}
\bibliography{acmart.bib}
\end{document}